\documentclass[preprint,showpacs,preprintnumbers,amsmath,amssymb]{revtex4-1}
\usepackage[dvipdfmx]{graphicx}
\usepackage{txfonts}
\begin{document}
\title{The Jellium Edge and the Size Effect of the Chemical Potential and Surface Energy in Metal Slabs}

\author{Kazuhiko Seki\thanks{k-seki@aist.go.jp}}
\affiliation
%\inst
{Nanomaterials Research Institute(NMRI), 
National Institute of Advanced Industrial Science and Technology (AIST)\\
AIST Tsukuba Central 5, Higashi 1-1-1, Tsukuba, Ibaraki, Japan, 305-8565, Japan} %\\

%\abst{
\begin{abstract}
Although free electron models have been established in order to capture the essential physics of interfacial 
and bulk properties in metals, 
some issues still remain regarding the application of free electron models to thin metal films. 
One of the issues relates to whether 
the geometric edge coincides with the potential edge in order to satisfy the charge neutrality condition when 
the potential profile is modeled as a rectangular potential well. 
We show that they coincide 
by rigorously taking into account 
 the quantization effect arising from electron confinement in a thin metal slab. 
  As a result, the overall behaviors of the chemical potential and surface energy show  
an increasing trend by decreasing the thickness of the slab. 
The chemical potential and surface energy show an oscillatory thickness dependence by further taking into account the discreteness of the total number of free electrons.
%}
\end{abstract}

%%% Keywords are not needed any longer. %%%
%%%\kword{keyword1, keyword2, keyword3, \ldots}
%%%

%\begin{document}
\maketitle

\section{Introduction}

Understanding quantum effects in nano-structured metals is important for developing nano-devices. 
Although free electron models have been used to capture the essential physics of interfacial 
and bulk properties in metals, \cite{kittel_80,Lang_70}
free electron models for thin metal films have not yet been fully investigated. 
In thin metal films, 
the separation of quantum states and maximum states occupied by electrons 
depend on the film thickness and should be evaluated with care.
One of the issues in free electron models for thin metal films relates to whether 
the geometric edge coincides with the potential edge in order to satisfy the charge neutrality condition when 
the potential profile is modeled as a rectangular potential well. 
Such details in this model affect the quantum effect in thin metal films. \cite{STRATTON_65,Kiejna_96,Kostrobij_18,SCHULTE_76}

It is necessary to consider the background positive charge and electron density 
to correctly determine the electrostatic potential in metals. \cite{Bardeen_36,Lang_70,Kiejna_96}
The simplest model is the so-called jellium model,  
wherein the background positive charge is expressed as a uniform profile in the bulk of 
the metal sample and is sharply terminated at the surface. 
The cut-off edge is called the jellium edge or the geometric edge.
The jellium edge is determined by imposing the charge neutrality condition, 
where the electron density is calculated quantum mechanically. 
It has been frequently stated that the jellium edge differs from the potential edge of the rectangular potential 
of free electrons, and others have concluded that 
the jellium edge and potential edge are shifted relative to each other. \cite{Bardeen_36,vanHimbergen_78,STRATTON_65,Kiejna_96,Huntington_51,Kostrobij_15,Kostrobij_18,Han_09,SCHULTE_76} 
On the other hand, the shift in the jellium edge away from the potential edge is not considered when calculating the chemical potential of metal films in some cases. \cite{Dymnikov_11,Paskin_65,SMITH_65,THOMPSON_63,Biao_08}
It has been argued whether or not the jellium edge is different from the potential surface. \cite{STRATTON_65,Kiejna_96,Kostrobij_18}
In this manuscript, we scrutinize whether 
the jellium edge coincides with or differs from the potential edge of the rectangular potential 
for free electrons. 

If the Fermi sphere is considered in the ground state (at temperature $T=0$ K), 
the number of free electrons in the volume $L^3$ is expressed by \cite{kittel_80}
\begin{align}
N=2 \frac{(4\pi/3)K_F^3}{(2\pi/L)^3}.
\label{eq:FS}
\end{align}
When the number of positive charges from ions is equal to the number of free electrons, 
the average density of positive charge is  
\begin{align}
\rho=\frac{N}{L^3}=\frac{K_F^3}{3\pi^2}, 
\label{eq:FSr}
\end{align}
%%% revised
where $K_F$ denotes the bulk Fermi wave vector. 
%%% revised
The average positive charge density given by Eq. (\ref{eq:FSr}) is a consequence of 
the Sommerfeld model for free electrons in an infinite box potential. 
For the infinite barrier model, 
the number of positive charges increases as $L^3$ in such a way that $N/L^3$ approaches 
a constant value given by Eq. (\ref{eq:FSr}). 

A free electron model of a finite width was introduced by imposing the Born-von Karman periodic boundary condition in the x and y directions, as well as a fixed boundary condition at $z=L_z$ without taking the thermodynamic limit. \cite{Bardeen_36}
The electron density profile along the z-direction
shows oscillation with a wavelength given by $\pi/K_F$,
called Friedel oscillations. \cite{Kiejna_96}
Friedel oscillations will be discussed later for some particular cases. 

On the basis of the electron density, 
the jellium edge is obtained by applying the charge neutrality condition, where   
the average positive charge density should be correctly evaluated. In this paper, we rigorously evaluate  
the average positive charge density by taking into account the quantum effect arising from 
the finite width of the metal slab. 
By using the charge neutrality condition together with the rigorous results regarding the average positive charge density and the electron density profile, 
we show that the jellium edge coincides with the potential edge. 
On this basis, 
we study the size dependence of the chemical potential and the surface energy due to 
quantum confinement of free electrons in a slab. 

%%%%%%%%%%%%%%%%%%%%%%%%%%%%%%%%%%%%%%%%%%%%%%%%%
\section{Theory}
We consider 
a slab of finite thickness $L_z$ in the z direction. 
The slab is extended infinitely in the other directions and, 
the Born-von Karman periodic boundary condition is imposed along the x and y directions with characteristic lengths $L_x$ and $L_y$, respectively. 
For simplicity, we first consider the case where 
the potential is zero at $0\leq z \leq L_z$ and is 
infinitely large at $z=0$ and $z=L_z$.
The wave function inside the slab is expressed as \cite{vanHimbergen_78,Kiejna_96}
\begin{align}
\psi_k (x,y,z)= \sqrt{\frac{2}{L_x L_y L_z}} \exp 
\left[ i(k_x x+ k_y y) \right] \sin (k_z z ) ,
\label{eq:wf_i}
\end{align}
where the components of the wave vectors are given by 
\begin{align}
k_x= \pm \frac{2\pi}{L_x} n_x, k_y=\pm  \frac{2\pi}{L_y} n_y \mbox{ } n_{x,y}=0,1, 2, \cdots 
\label{eq:kykz}\\
k_z= \frac{\pi}{L_z} n_z, \mbox{ } n_{z}=1, 2, 3, \cdots . 
\label{eq:kx}
\end{align}
%%% changed 
In the ground state (at temperature $T=0$ K), 
$n_x$, $n_y$, and $n_z$ satisfy 
$\hbar^2 (k_x^2 + k_y^2 + k_z^2)/(2m) \leq \hbar^2 k_F^2/(2m)$, 
where $m$ and $\hbar$ are the electron mass and 
the Planck constant divided by
$2 \pi$, respectively.  
$k_F$ indicates the Fermi wave vector, {\it i.e.}, the largest wave vector occupied by electrons in the ground state. 
As shown later, $k_F$ depends on the thickness $L_z$ and should be distinguished from 
the bulk Fermi wave vector denoted by $K_F$ except when the limit of $L_z \rightarrow \infty$ is taken. 
%%% changed 

The number of allowed values for $k_x$ and $k_y$ for a certain value of $k_z$ is given by 
\begin{align}
\frac{L_x L_y}{(2\pi)^2} \pi (k_F^2-k_z^2)=\frac{1}{4 \pi} L_x L_y \left(k_F^2-k_z^2 \right).
\label{eq:allowed}
\end{align}
The electron density profile in the ground state is obtained 
as 
%%% changed 
\cite{vanHimbergen_78,Sugiyama_60}
\begin{align}
n_e (z)&= 2 \frac{1}{4 \pi} L_x L_y\sum_{n_z=1}^{n_{\rm Fz}} \left(k_F^2-k_z^2 \right)\left|\psi_k (x,y,z) \right|^2\\
&= \frac{1}{2\pi L_z} \sum_{n_z=1}^{n_{\rm Fz}} \left(k_F^2 - k_z^2 \right)
\left[1- \cos \left(2 k_z z \right) \right] ,
\label{eq:edensity_i}
\end{align}
where spin multiplicity is included. 
%%% changed 
In Eq. (\ref{eq:edensity_i}), 
$n_{\rm Fz}$ is the quantum number for the highest occupied $n_{z}$ and 
is given by 
$n_{\rm Fz}=[k_F L_z/\pi] 
\approx k_F L_z/\pi$, 
where $[x]$ indicates the integer part of $x$. 
%%% changed 
The summation in Eq. (\ref{eq:edensity_i}) with $k_z=(\pi /L_z)n_z$ is evaluated analytically and is simplified 
by using Mathematica \cite{Mathematica};  
the electron density can be expressed as  
\begin{multline}
n_e (z)= 
\frac{k_F^2}{2 \pi L_z}\left(n_{\rm Fz}+\frac{1}{2} \right)- \frac{\pi}{12 L_z^3} n_{\rm Fz} (n_{\rm Fz}+1)(2n_{\rm Fz}+1) 
\\
+\frac{1}{8\pi L_z^3} \left[(2n_{\rm Fz}+3) \pi^2
\cos\left(\frac{2(n_{\rm Fz}+1)\pi z}{L_z} \right)
\csc^2 \left(\frac{\pi z}{L_z} \right)
\right. 
\\
 + 2 
\left( \pi^2 (n_{\rm Fz}+1)^2 -k_F^2 L_z^2\right) 
 \sin\left(\frac{(2n_{\rm Fz}+1)\pi z}{L_z} \right)
\csc \left(\frac{\pi z}{L_z} \right) 
%\\
\left.
 - \pi^2 \sin\left(\frac{(2n_{\rm Fz}+3)\pi z}{L_z} \right)
\csc^3 \left(\frac{\pi z}{L_z} \right) 
 \right] . 
\label{eq:edensity_i2}
\end{multline}
The total number of electrons is calculated using
\begin{align}
\int_0^{L_z} n_e (z) dz &= \frac{1}{2\pi L_z} \sum_{n_z=1}^{n_{\rm Fz}} \left(k_F^2 - k_z^2 \right)
\int_0^{L_x} dz \left[1-\cos
\left(2 \frac{\pi}{L_z} n_z z \right)\right] \nonumber \\
&= \frac{1}{2\pi L_z^2} \sum_{n_z=1}^{n_{\rm Fz}} \left(k_F^2 L_z^2 - n_z^2 \pi^2 \right)\\
&= \frac{k_F^2 n_{\rm Fz}}{2\pi}  - \frac{\pi}{12 L_z^2}  n_{\rm Fz}(n_{\rm Fz}+1)(2n_{\rm Fz}+1).  
\label{eq:Telectrons_i}
\end{align}
Equation (\ref{eq:Telectrons_i}) is known. \cite{THOMPSON_63}
%Direct integration of Eq. (\ref{eq:edensity_i2}) with Mathematica yields a lengthy expression. \cite{Mathematica}
%We confirm that the lengthy expression reduces to Eq. (\ref{eq:Telectrons_i}) for $n_{\rm Fz}=1,2,\cdots, 5$ by using Mathematica. \cite{Mathematica}

The total number of positive charges from ions 
is equal to the total number of free electrons denoted by $N$.
The total number of positive charges $N$ can be calculated by calculating a relation for the total number of electrons, given by 
\begin{align}
2 \frac{1}{4\pi} L_x L_y \sum_{n_z=1}^{n_{\rm Fz}} \left(k_F^2  - k_z^2 \right)=N, 
\label{eq:Tions_i}
\end{align}
which can be rewritten as \cite{Czoschke_05}
\begin{align}
\frac{L_x L_y n_{\rm Fz}}{2 \pi}\left[k_F^2 -\frac{1}{6 L_z^2} (n_{\rm Fz}+1)(2n_{\rm Fz}+1)\pi^2\right]=N. 
\label{eq:N_i} 
\end{align}
We find from Eq. (\ref{eq:N_i}) that
\begin{align}
k_F^2=\frac{\pi^2}{6 L_z^2} (n_{\rm Fz}+1)(2n_{\rm Fz}+1) + \frac{2\pi N}{L_x L_y n_{\rm Fz}} . 
\label{eq:kF_i}
\end{align} 
By substituting Eq. (\ref{eq:kF_i}) into Eq. (\ref{eq:Telectrons_i}), we obtain 
\begin{align}
\frac{1}{L_z} \int_0^{L_z} dz n_e (z) = \frac{N}{L_x L_y L_z}. 
\label{eq:av_i}
\end{align}
The average positive charge density (denoted by $\rho$) should be given by 
the number of positive charges $N$ averaged over the volume (given by $L_x L_y L_z$), {\it i.e.},
$\rho=N/(L_x L_y L_z)$.  
Equation (\ref{eq:av_i}) indicates that charge neutrality is satisfied; 
it is not necessary to introduce a new width for the ions instead of $L_z$ to describe the shift of the jellium edge from the potential edge.
Therefore, the jellium edge coincides with the potential edge.

The shift of the jellium edge from the potential edge follows from the following argument. 
If we loosely evaluate Eq. (\ref{eq:edensity_i}) by integration, we have \cite{Kiejna_96,Sugiyama_60}
\begin{align}
n_e^{(a)} (z)&=\frac{L_z}{\pi}\frac{1}{2\pi L_z} \sum_{n_z=1}^{n_{\rm Fz}} \left(k_F^2 - k_z^2 \right)
\left[1- \cos \left(2 k_z z \right) \right] \frac{\pi}{L_z} \\
&\approx \frac{1}{2\pi^2} \int_0^{K_F} d k_z 
\left(K_F^2 - k_z^2 \right)\left[1- \cos \left(2 k_z z \right) \right] \\
&= \rho \left( 1+ \frac{3 \cos (2 K_F z)}{(2 K_F z)^2} - \frac{3 \sin (2 K_F z)}{(2 K_F z)^3}
\right),
\label{eq:loose}
\end{align}
%%% revised 
where $K_F$ appeared by evaluating $k_F$ in the limit of $L_z \rightarrow \infty$, and $\rho=K_F^3/(3\pi^2)$ is given by 
Eq. (\ref{eq:FSr}).  
%%% revised 
Similarly, we can loosely evaluate the average positive charge density by integration using Eq. (\ref{eq:Tions_i}): \cite{Kiejna_96,Sugiyama_60}
\begin{align}
 \frac{1}{2\pi^2}L_x L_y L_z \int_0^{K_F}d k_z \left(K_F^2  - k_z^2 \right)=
 \frac{K_F^3}{3\pi^2} L_x L_y L_z =N. 
\label{eq:Tions_ii}
\end{align}
Therefore, the average positive charge density is given by $N/(L_x L_y L_z)$ and is obtained as $\rho$. 
We also find from Eq. (\ref{eq:loose}) 
\begin{align}
\int_0^\infty dz \left[n_e^{(a)} (z) - \rho\right]=- \rho \frac{3\pi}{8K_F}.
\label{eq:shift_i}
\end{align}
If one notices that a single edge is present in the limit of $\pi/L_z \rightarrow 0$, the result suggests that 
the jellium edge is $3\pi/(8K_F)$ and is shifted inside from the potential edge to satisfy the charge neutrality condition;  \cite{Bardeen_36,vanHimbergen_78,Kiejna_96,Huntington_51,Kostrobij_15,Han_09,SCHULTE_76}
Equation (\ref{eq:shift_i}) indicates 
$\int_0^{L_z} dz n_e^{(a)} (z) =\rho[L_z- (3\pi)/(8K_F)]$ 
when $L_z$ is sufficiently larger than the Wigner-Seitz radius. 
The difference between the density calculated from the wave function and 
that calculated from the density of states is a mathematical consequence of 
taking the limit $\pi/L_z \rightarrow 0$. 
The difference disappears as long as we rigorously evaluate the quantized effect represented by summation, 
thus the difference is unphysical in thin slabs. 
We conclude that the jellium edge coincides with the potential edge for free electrons 
at least when the potential barrier along the z-direction is symmetric and infinite. 
Below, we show that the same conclusion can be reached for more general potentials. 

We consider 
a slab of finite thickness $L_z$ in the z-direction and extended infinitely in the x and y directions, where  
the Born-von Karman periodic boundary condition is imposed in x and y directions as before. 
The potential is zero at $0\leq z \leq L_z$. 
A constant potential denoted by $w_R$ 
is assumed for $L_z \leq z$, and 
a constant potential denoted by $w_L$ is assumed for $z\leq 0$. 
The values of both $w_R$ and $w_L$ are assumed to be positive. 
The wave function is denoted by $\psi_n (z)\exp 
\left[ i(k_x x+ k_y y) \right]/\sqrt{L_x L_y}$, 
and the eigen-states are denoted by $n=1,2,\cdots$.
Because the eigen-states are orthonormal, 
the electron density can be written as 
\begin{align}
\int_{-\infty}^\infty dz n_e (z)&= \frac{1}{2\pi} \sum_{n_z=1}^{n_{\rm Fz}} \left(k_F^2 - k_z^2 \right) 
\int_{-\infty}^\infty dz\left| \psi_n (z) \right|^2 
\nonumber \\
&=\frac{1}{2\pi} \sum_{n_z=1}^{n_{\rm Fz}} \left(k_F^2 - k_z^2 \right) . 
\label{eq:orthonormalized}
\end{align}
The total number of positive charges obeys Eq. (\ref{eq:Tions_i}) in the ground state. 
By combining Eq. (\ref{eq:orthonormalized}) and Eq. (\ref{eq:Tions_i}), 
we obtain,  
\begin{align}
\frac{1}{L_z} \int_{-\infty}^\infty dz n_e (z) = \frac{N}{L_x L_y L_z}. 
\label{eq:density_f}
\end{align}
Equation (\ref{eq:density_f}) expresses the charge neutrality condition, where the positive charge density
is obtained by averaging the total number of positive charges over the potential width denoted by $L_z$, yielding $\rho=N/(L_x L_y L_z)$. 
If the jellium edge differs from the potential edge, then $L_z$ on the left-hand side of 
Eq. (\ref{eq:density_f}) must be changed. 
Therefore, we conclude that the jellium edge coincides with the potential edge.

%%%%%%%%%%%%%%%%%%%%%%%%%%%%%%%%%%%%%%%%%%%%%%
\section{Chemical potential results}

By substituting the positive charge density given by $\rho=N/(L_x L_y L_z)$ into Eq. (\ref{eq:kF_i}) 
and using the Wigner-Seitz radius given by $r_s=[3/(4\pi \rho)]^{1/3}$,
we obtain a closed equation giving the thickness dependence of the Fermi wave vector as 
\begin{align}
k_F^2=
\frac{\pi^2}{6 L_z^2} 
 (n_{\rm Fz}+1)(2n_{\rm Fz}+1) + \frac{3}{2 n_{\rm Fz}}  \frac{L_z}{r_s^3}.\,  
\label{eq:closedkF}
\end{align}
where $n_{\rm Fz}=[k_F L_z/\pi] 
\approx k_F L_z/\pi$. 
By introducing the approximation given by $n_{\rm Fz}\approx k_F L_z/\pi$, 
we find 
\begin{align}
k_F &= \frac{\pi}{4 L_z} \left[1+ \frac{7 \pi^{2/3}}{f(\ell_z)^{1/3}} + 
\frac{f(\ell_z)^{1/3}}{3\pi^{2/3}} 
\right]
\label{eq:kFsole}\\
&\approx \left(\frac{9}{4} \pi \right)^{1/3} \frac{1}{r_s} + \frac{\pi}{4 L_z} + 
\frac{7 \pi^{5/3} r_s}{24 \times 18^{1/3} L_z^2} ,
\label{eq:kFsola}
\end{align}
where $f(\ell_z)= 2^3 \times 3^5 \ell_z^3 + 81 \pi^2 + 6 
\sqrt{16 \times 3^8 \ell_z^6 + 4 \times 3^7 \ell_z^3 \pi^2 - 75 \pi^4
}
$ and $\ell_z=L_z/r_s$.
Apart from a constant in terms of $L_z$, 
the chemical potential of free electrons can be expressed as $\mu=\hbar^2 k_F^2/(2m)$. 
Consequently, the size dependence of the chemical potential can be rewritten as 
\begin{align}
\mu\approx \frac{\hbar^2 }{2m}
\left[
\left(\frac{9}{4} \pi \right)^{1/3} \frac{1}{r_s} + \frac{\pi}{4 L_z} + 
\frac{7 \pi^{5/3} r_s}{24 \times 18^{1/3} L_z^2}  \right]^2. 
\label{eq:chempo1}
\end{align}
In the limit $L_z \rightarrow \infty$, 
we recover 
the bulk chemical potential expressed as \cite{kittel_80}
\begin{align}
\mu_b=\frac{\hbar^2 }{2m} \left(3 \pi^2 \rho \right)^{2/3}
=\frac{\hbar^2 }{2m r_s^2} \left(\frac{9 \pi}{4} \right)^{2/3} . 
\label{eq:bulkmu}
\end{align} 

\begin{figure}[h]
  \begin{center}
    \includegraphics[width=0.3\textwidth]{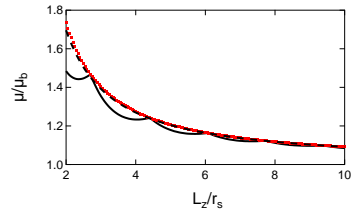}
  \end{center}
  \caption{(Color online) Normalized chemical potential is shown against the thickness of the slab. 
    The thickness of the slab is normalized by the Wigner-Seitz radius given by 
    $r_s=[3/(4\pi \rho)]^{1/3}$. 
  The dashed line is calculated using Eq. (\ref{eq:kFsole}) and normalized by the bulk value given by $\mu_b$
  [Eq. (\ref{eq:bulkmu})]. 
    The (red) dots are calculated from Eq. (\ref{eq:chempo1}). 
  The solid line indicates the exact result shown in the main text. 
}
  \label{fig:Chempot}
\end{figure}

In Fig \ref{fig:Chempot}, we show the chemical potential $\hbar^2 k_F^2/(2m)$ of free electrons as a function of $L_z$ 
calculated from Eq. (\ref{eq:chempo1}).
We also show the exact chemical potential obtained using Eq. (\ref{eq:kF_i}) and $n_{\rm Fz}=[k_F L_z/\pi]$. 
%%% changed 
The figure is invariant under changes in the electron density, which indicates 
that the free electron model is characterized by a single length scale given by the Wigner-Seitz radius.  
Equation (\ref{eq:chempo1}) indicates that the chemical potential is a monotonic decreasing function of $L_z$. 
The separation between quantum states increases by decreasing the thickness of the slab; 
the largest energy value of electrons 
in the ground state ($\mu$)
increases by increasing the separation between quantum states. 
This qualitative feature was already pointed out previously. \cite{Czoschke_05,Korotun_15} 
The exact numerical result in Fig. \ref{fig:Chempot} shows the oscillatory dependence on $L_z$.
The oscillatory $L_z$ dependence originates from the discreteness of the total number of free electrons. 
This result agrees with the oscillatory $L_z$ dependence of the chemical potential reported previously.  \cite{Czoschke_05,Korotun_15,Kostrobij_15,Kostrobij_18}

\begin{figure}[h]
  \begin{center}
    \includegraphics[width=0.3\textwidth]{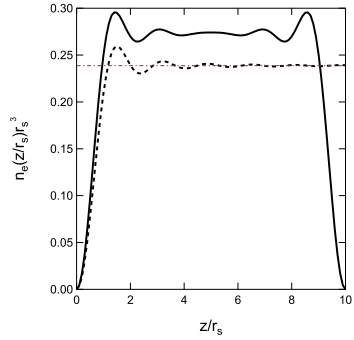}
  \end{center}
  \caption{(Color online)  The electron density is shown as a function of distance from the surface when 
  $L_z=10 r_s$.  The electron density is made dimensionless by multiplying it with $r_s^3$.
  Both the electron density and the distance from the surface of the slab are normalized by the Wigner-Seitz radius $r_s=[3/(4\pi \rho)]^{1/3}$. 
  The solid line shows the exact result obtained from Eq. (\ref{eq:edensity_i2}). 
  The normalized average density is 
  shown by the (red) dashed-dotted line, which is  
  given by $3/(4\pi)$ because $\rho=3/(4\pi r_s^3)$. 
The dashed line shows the result from Eq. (\ref{eq:loose}) 
obtained in the limit $\pi/L_z \rightarrow 0$.
}
  \label{fig:density}
\end{figure}

In Fig. \ref{fig:density}, 
we show the electron density profile given by Eq. (\ref{eq:edensity_i2}) and a comparison with the electron density obtained from Eq. (\ref{eq:loose}), where the limit $\pi/L_z \rightarrow 0$ is employed. 
The electron density is made dimensionless by multiplying it with $r_s^3$. 
The figure is invariant under changes in the electron density.  
The result in Eq. (\ref{eq:edensity_i2}) 
indicates that the electron density satisfies the boundary condition at $z=0$ and $z=L_z$. 
The electron density in the middle of the thickness is higher than the average density 
because of the depletion of the electron density near the boundaries due to quantum confinement effect. 
By increasing the slab thickness, 
the difference between the average density and the density in the middle of the thickness decreases. 
%%% revised 
When the positive charge density with the average electron density is uniformly distributed over the thickness determined by 
the potential edges, 
the electron density  in the middle of the slab is higher than the positive charge density. 
If the positive charge density is assumed to be equal to the electron density in the middle of 
the thickness, 
the positive charge density should be distributed over the thickness narrower than 
that determined from the potential edges to maintain the overall charge neutrality in the slab. 
In this case, the positive charge density differs from the average electron density;  
as a result, the geometric edges differ from the potential edges. 
Both types of edges coincide when the positive charge density with the average electron density is 
uniformly distributed. 
%%% revised 

The electron density profile obtained from Eq. (\ref{eq:edensity_i2}) 
shows oscillation with wavelength $L_z/n_{\rm Fz}$. 
Using $n_{\rm Fz}=[k_F L_z/\pi] 
\approx k_F L_z/\pi$, 
and 
Eq. (\ref{eq:kFsole}) for $k_F$,
the wavelength can be written as 
\begin{align}
\frac{L_z}{n_{\rm Fz}}\approx k_F/\pi \approx\left( \frac{2\pi}{3} \right)^{2/3} r_s \approx 1.64 r_s. 
\label{eq:FO1}
\end{align}
The oscillation shown in Fig. \ref{fig:density} is well characterized by the wavelength. 
If the bulk value of the Fermi wave vector obtained from Eq. (\ref{eq:FSr}) is introduced into Eq. (\ref{eq:FO1}), 
%%% revised 
we find $K_F/\pi$.
The result in Eq. (\ref{eq:loose}) also shows oscillation with a wavelength given by $K_F/\pi$.
%%% revised
However, it 
differs significantly from the exact result. 
The density of electrons obtained in the limit $\pi/L_z \rightarrow 0$ is lower than 
the exact result in the middle of the slab thickness. 
Moreover, we see that the approximate electron density does not fulfill the boundary condition at $z=L_z$.
%%% revised
The electron density approaches zero only at $z=0$. 
%%% revised

Before closing this section, we comment on the energy required to excite an electron from the ground state. 
If we denote the right-hand side of Eq. (\ref{eq:kF_i}) by $k_F^2 (n_{\rm Fz})$,  
the excitation energy can be estimated from 
$\Delta \epsilon_F = \hbar^2/(2m)[k_F^2 (n_{\rm Fz}+1)-k_F^2 (n_{\rm Fz})]$ as  
\begin{align}
\Delta  \epsilon_F &= \frac{\hbar^2 }{12m L_z^2} \left(
\pi^2 \left( 5+4 n_{\rm Fz} \right) - \frac{9 (L_z/r_s)^3}{n_{\rm Fz}(n_{\rm Fz}+1)} \right) \\
&=\frac{\pi^2\hbar^2 }{m L_z^2} \left(
1 - \frac{1}{3} \frac{K_F^2 r_s^3}{L_z}  \right),
\label{eq:Kubo1}
\end{align}
where Eq. (\ref{eq:kFsola}) is used. 
For sodium, $\Delta  \epsilon_F $ is estimated to be about $0.026$ (eV) when 
the slab thickness is $5.3$ (nm) using Eq. (\ref{eq:Kubo1}). 
If the Wigner-Seitz radius is given by $r_s=3$ (Bohr), 
 $\Delta  \epsilon_F $ is also about $0.026$ (eV) when 
the slab thickness is $5.3$ (nm). 
The results suggest the length scale of 
the Kubo effect at room temperature; the quantization of one electronic level at the Fermi level 
results in remarkable effects in thermodynamic properties of fine metals. \cite{Kubo_62}
Equation (\ref{eq:Kubo1}) indicates that the Kubo effect 
can be observed for the thicker metal slabs if the temperature is lowered from room temperature.

%%%%%%%%%%%%%%%%%%%%%%%%%%%%%%%%%%%%%%%%%%%%%%
\section{Surface energy results}

Similarly, we can calculate the surface energy. 
First, we express the total energy as  \cite{Czoschke_05}
\begin{align}
E_t&= \frac{\hbar^2}{2m}  \frac{2L_x L_y}{(2\pi)^2} \sum_{n_z=1}^{n_{\rm Fz}}  \int_0^{\sqrt{k_F^2 - k_z^2}} d k_{\parallel} \,
2 \pi k_{\parallel} \left(k_{\parallel}^2+k_z^2 \right)\\
&= \frac{\hbar^2}{2m}  \frac{L_x L_y}{4\pi} \sum_{n_z=1}^{n_{\rm Fz}}  \left(k_F^4 - k_z^4 \right) \\
&= \frac{\hbar^2}{2m}  \frac{L_x L_y}{4\pi} 
\left[n_{\rm Fz} k_F^4- \frac{\pi^4}{30 L_z^4} n_{\rm Fz}(n_{\rm Fz}+1)(2n_{\rm Fz}+1)(3n_{\rm Fz}^2+3n_{\rm Fz}-1)
\right], 
\label{eq:Et}
\end{align}
where $k_z$ is given by Eq. (\ref{eq:kx}). 
Then, we decompose the total energy into the bulk energy per unit volume and the two parts of the surface energy per unit area of the slab as \cite{Czoschke_05}
\begin{align}
E_t&= \epsilon_b L_x L_y L_z + 2 \epsilon_s L_x L_y ,
\label{eq:Etd}
\end{align} 
where the bulk energy is obtained 
using $n_{\rm Fz} \approx k_F L_z/\pi$ 
as 
\begin{align}
\epsilon_b&=\lim_{L_z \rightarrow \infty} \frac{E_t}{L_x L_y L_z}\\
&= \frac{\hbar^2 K_F^5}{10m \pi^2} ,
\label{eq:eb}
\end{align}
where $k_F$ is replaced with $K_F$ since 
the bulk energy can be obtained in the limit of $L_z \rightarrow \infty$. 
The obtained bulk energy is equal to 
the total energy given by $(3 N/5) \hbar^2 K_F^2/(2m)$ divided by $L_x L_y L_z$ 
as it should be.  \cite{kittel_80} 
Now, we calculate the surface energy. 
The surface energy is obtained 
from Eq. (\ref{eq:Et})  
using the decomposition given by Eq. (\ref{eq:Etd})  
as 
\begin{align}
\frac{\epsilon_s}{\epsilon_{s0}}&= 
2 \left( \frac{4}{9\pi}\right)^{4/3} 
\left[
n_{\rm Fz} (k_F r_s)^4- 
\frac{\pi^4 r_s^4}{30 L_z^4} n_{\rm Fz}(n_{\rm Fz}+1)(2n_{\rm Fz}+1)(3n_{\rm Fz}^2+3n_{\rm Fz}-1)
\right]-
\frac{8}{5 \pi} K_F L_z
\label{eq:es}\\
&\approx 1 + \frac{\pi}{2 K_F L_z} - \frac{\pi^2}{24 (K_F L_z)^2} ,
\label{eq:esa}
\end{align}
where the surface energy in the limit of $L_z \rightarrow \infty$ is given by 
\begin{align}
\epsilon_{s0}&=\frac{1}{32 \pi} \frac{\hbar^2 K_F^4}{m} .
\label{eq:es0}
\end{align}
Equation (\ref{eq:es0}) was obtained earlier. \cite{waber_72}

\begin{figure}[h]
  \begin{center}
    \includegraphics[width=0.3\textwidth]{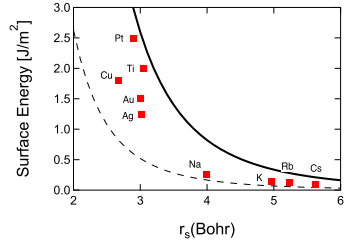}
  \end{center}
  \caption{(Color online) The surface energy is shown against the Wigner-Seitz radius.  
  The solid line is obtained from Eq. (\ref{eq:es0}) and the dashed line indicates the classical result of 
  $\hbar^2 k_F^4/(160 \pi m)$. 
The (red) dots indicate the experimental values of the surface energies of metals.
}
  \label{fig:SF}
\end{figure}

In Fig. \ref{fig:SF}, we compare the surface energy given by Eq. (\ref{eq:es0}) 
with the experimental values. 
The metal species are chosen so that 
the Sommerfeld parameter of the heat capacity is close to that estimated from the free electron model. 
Specifically, the criterion is that 
the ratio of the measured to the free electron values of the Sommerfeld parameter is between 
$0.7$ and $1.3$. 
The solid line is the surface energy calculated from the free electron model of the slab 
obtained from Eq. (\ref{eq:es0}). 
Compared to the Sommerfeld parameter, 
the larger deviation of the experimental values from the theoretical results of the free electron model 
is found for 
the surface energy  of the slab. 
The deviation could originate from the oversimplification in the free electron model of the slab, 
where the inhomogeneity in the background positive charges in particular near the surface is ignored.  
Though we did not consider finite barriers, 
the effect of finite barrier hight  
can be taken into account in a straightforward manner as sketched in Theory section and 
can be ignored as long as the lowest barrier hight sufficiently exceeds the chemical potential.  
 In the same figure, we show the known result given by 
 %%% revised 
 $\epsilon_{s0} \approx \hbar^2 K_F^4/(160 \pi m)$ which is derived 
 %%% revised
 by applying the charge neutrality condition
using an average positive charge density that is different 
from %%% changed
the expression presented here. \cite{Huntington_51,Sugiyama_60} 
We consider the free electron model of the slab; 
the surface energy 
is calculated from the natural decomposition of the total energy into the bulk part and the surface parts. 
Then,  
the summation appeared due to quantization in the direction of the slab thickness is 
rigorously evaluated. 
The deviation introduced by approximating the summation by integration is well captured in 
the electron density shown in Fig. \ref{fig:density}.
As a result, the dashed line significantly differs from the solid line in Fig. \ref{fig:SF}.

\begin{figure}[h]
  \begin{center}
    \includegraphics[width=0.3 \textwidth]{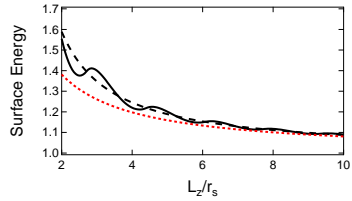}
  \end{center}
  \caption{(Color online) Normalized surface energy is shown against the thickness of the slab. 
    The thickness of the slab is normalized by the Wigner-Seitz radius given by 
    $r_s=[3/(4\pi \rho)]^{1/3}$. 
  The dashed line is calculated from Eq. (\ref{eq:es}) with 
  $n_{\rm Fz} 
\approx k_F L_z/\pi$
  and normalized by the bulk value of the surface energy
  [Eq. (\ref{eq:es0})]. 
  The (red) dots are calculated from Eq. (\ref{eq:esa}). 
  The solid line indicates the exact result obtained from Eq. (\ref{eq:es}) and Eq. (\ref{eq:closedkF})
  using $n_{\rm Fz}=[k_F L_z/\pi]$. 
}
  \label{fig:nSF}
\end{figure}

In Fig. \ref{fig:nSF}, 
we show the surface energy as a function of $L_z$ 
calculated from Eq. (\ref{eq:es}). 
The surface energy is normalized by that taking the limit $L_z \rightarrow \infty$. 
We also show the surface energy by taking into account the discreteness of the total number of 
free electrons. 
By substituting $k_F$ obtained from Eq. (\ref{eq:closedkF}) into Eq. (\ref{eq:es}) 
using $n_{\rm Fz}=[k_F L_z/\pi]$, 
we obtained the exact result of the surface energy.
The figure is invariant under changes in the electron density. 
Equation (\ref{eq:esa}) indicates that the surface energy is a monotonic decreasing function of $L_z$ like  
the chemical potential. 
The exact numerical result in Fig. \ref{fig:nSF} shows the oscillatory dependence on $L_z$.
The oscillatory $L_z$ dependence originates from the discreteness of the total number of free electrons 
as in the case of the chemical potential. 

%%%%%%%%%%%%%%%%%%%%%%%%%%%%%%%%%%%%%%%%%%%%
\section{Conclusion}

We have considered the electron density and the number of positive charges by rigorously taking into account 
the quantum effect while keeping the discrete sum in the free electron model for a thin metal slab. 
We showed that the jellium edge coincides with the rectangular potential edge. 
The effect of quantum confinement on the chemical potential %%% changed 
and surface energy in a thin slab was subsequently studied. 
The thickness dependence of the chemical potential was derived as Eq. (\ref{eq:chempo1});  
the chemical potential (Fermi energy) increases 
as the thickness of the slab decreases because the separation between quantized states becomes wider. 
The thickness dependence of the surface energy was derived as Eq. (\ref{eq:esa}) 
and showed the similar dependence on the thickness of the slab.
More accurate results for the chemical potential and surface energy 
that reflect the discrete nature of the number of electrons in the Wigner-Seitz cell showed the oscillatory thickness dependence superimposed on top of the 
continuous thickness dependence mentioned above. 

In our analysis, quantized effects due to confinement of electrons in a thin slab were   
considered according to the free electron model, where the length scale is characterized by the Wigner-Seitz radius.
Lattice structures and lattice constants could affect the physical quantities as the slab thickness decreases. 
A full quantitative characterization of a particular metal film 
based on the aforementioned structure is beyond the scope of 
the present study. 
The electrostatic interaction, the exchange interaction and the correlation interaction were also ignored.
Nevertheless, we qualitatively discussed the quantum size effect on the chemical potential 
and the surface energy in a thin metal slab.
In some theories, \cite{vanHimbergen_78,Kiejna_96,Kostrobij_15,Kostrobij_18,Han_09,SCHULTE_76} 
the shift of the jellium edge from the potential edge was 
calculated, and the chemical potential was affected by the shift. 
We showed that such a shift is unnecessary if both the electron density and the total number of positive charges are evaluated 
by taking into account 
the finite width of the metal slab. 
By using the charge neutrality condition together with the 
rigorous results for the average positive charge density and the electron density profile, 
we showed that the jellium edge indeed coincides with the potential edge. 

%\begin{acknowledgment}

%\acknowledgment

%For environments for acknowledgement(s) are available: \verb|acknowledgment|, \verb|acknowledgments|, \verb|acknowledgement|, and \verb|acknowledgements|.

%\end{acknowledgment}

%\appendix
%\section{}
%\begin{thebibliography}{9}
%
%\bibliography{metalsurface_PR.bib}

\begin{thebibliography}{22}%
\makeatletter
\providecommand \@ifxundefined [1]{%
 \@ifx{#1\undefined}
}%
\providecommand \@ifnum [1]{%
 \ifnum #1\expandafter \@firstoftwo
 \else \expandafter \@secondoftwo
 \fi
}%
\providecommand \@ifx [1]{%
 \ifx #1\expandafter \@firstoftwo
 \else \expandafter \@secondoftwo
 \fi
}%
\providecommand \natexlab [1]{#1}%
\providecommand \enquote  [1]{``#1''}%
\providecommand \bibnamefont  [1]{#1}%
\providecommand \bibfnamefont [1]{#1}%
\providecommand \citenamefont [1]{#1}%
\providecommand \href@noop [0]{\@secondoftwo}%
\providecommand \href [0]{\begingroup \@sanitize@url \@href}%
\providecommand \@href[1]{\@@startlink{#1}\@@href}%
\providecommand \@@href[1]{\endgroup#1\@@endlink}%
\providecommand \@sanitize@url [0]{\catcode `\\12\catcode `\$12\catcode
  `\&12\catcode `\#12\catcode `\^12\catcode `\_12\catcode `\%12\relax}%
\providecommand \@@startlink[1]{}%
\providecommand \@@endlink[0]{}%
\providecommand \url  [0]{\begingroup\@sanitize@url \@url }%
\providecommand \@url [1]{\endgroup\@href {#1}{\urlprefix }}%
\providecommand \urlprefix  [0]{URL }%
\providecommand \Eprint [0]{\href }%
\providecommand \doibase [0]{http://dx.doi.org/}%
\providecommand \selectlanguage [0]{\@gobble}%
\providecommand \bibinfo  [0]{\@secondoftwo}%
\providecommand \bibfield  [0]{\@secondoftwo}%
\providecommand \translation [1]{[#1]}%
\providecommand \BibitemOpen [0]{}%
\providecommand \bibitemStop [0]{}%
\providecommand \bibitemNoStop [0]{.\EOS\space}%
\providecommand \EOS [0]{\spacefactor3000\relax}%
\providecommand \BibitemShut  [1]{\csname bibitem#1\endcsname}%
\let\auto@bib@innerbib\@empty
%</preamble>
\bibitem [{\citenamefont {Kittel}\ and\ \citenamefont
  {Kroemer}(1980)}]{kittel_80}%
  \BibitemOpen
  \bibfield  {author} {\bibinfo {author} {\bibfnamefont {C.}~\bibnamefont
  {Kittel}}\ and\ \bibinfo {author} {\bibfnamefont {H.}~\bibnamefont
  {Kroemer}},\ }\href@noop {} {\emph {\bibinfo {title} {Thermal Physics}}}\
  (\bibinfo  {publisher} {W. H. Freeman},\ \bibinfo {year} {1980})\BibitemShut
  {NoStop}%
\bibitem [{\citenamefont {Lang}\ and\ \citenamefont {Kohn}(1970)}]{Lang_70}%
  \BibitemOpen
  \bibfield  {author} {\bibinfo {author} {\bibfnamefont {N.~D.}\ \bibnamefont
  {Lang}}\ and\ \bibinfo {author} {\bibfnamefont {W.}~\bibnamefont {Kohn}},\
  }\href {\doibase 10.1103/PhysRevB.1.4555} {\bibfield  {journal} {\bibinfo
  {journal} {Phys. Rev. B}\ }\textbf {\bibinfo {volume} {1}},\ \bibinfo {pages}
  {4555} (\bibinfo {year} {1970})}\BibitemShut {NoStop}%
\bibitem [{\citenamefont {Stratton}(1965)}]{STRATTON_65}%
  \BibitemOpen
  \bibfield  {author} {\bibinfo {author} {\bibfnamefont {R.}~\bibnamefont
  {Stratton}},\ }\href {\doibase https://doi.org/10.1016/0031-9163(65)90775-4}
  {\bibfield  {journal} {\bibinfo  {journal} {Phys. Lett.}\ }\textbf {\bibinfo
  {volume} {19}},\ \bibinfo {pages} {556 } (\bibinfo {year}
  {1965})}\BibitemShut {NoStop}%
\bibitem [{\citenamefont {Kiejna}\ and\ \citenamefont
  {Wojciechowski}(1996)}]{Kiejna_96}%
  \BibitemOpen
  \bibfield  {author} {\bibinfo {author} {\bibfnamefont {A.}~\bibnamefont
  {Kiejna}}\ and\ \bibinfo {author} {\bibfnamefont {K.}~\bibnamefont
  {Wojciechowski}},\ }\href@noop {} {\emph {\bibinfo {title} {Metal Surface
  Electron Physics}}}\ (\bibinfo  {publisher} {Pergamon},\ \bibinfo {address}
  {Oxford},\ \bibinfo {year} {1996})\BibitemShut {NoStop}%
\bibitem [{\citenamefont {{Kostrobij}}\ and\ \citenamefont
  {{Markovych}}(2018)}]{Kostrobij_18}%
  \BibitemOpen
  \bibfield  {author} {\bibinfo {author} {\bibfnamefont {P.~P.}\ \bibnamefont
  {{Kostrobij}}}\ and\ \bibinfo {author} {\bibfnamefont {B.~M.}\ \bibnamefont
  {{Markovych}}},\ }\href@noop {} {\bibfield  {journal} {\bibinfo  {journal}
  {ArXiv e-prints}\ } (\bibinfo {year} {2018})},\ \Eprint
  {http://arxiv.org/abs/1804.08884} {arXiv:1804.08884 [cond-mat.stat-mech]}
  \BibitemShut {NoStop}%
\bibitem [{\citenamefont {Schulte}(1976)}]{SCHULTE_76}%
  \BibitemOpen
  \bibfield  {author} {\bibinfo {author} {\bibfnamefont {F.}~\bibnamefont
  {Schulte}},\ }\href {\doibase https://doi.org/10.1016/0039-6028(76)90250-8}
  {\bibfield  {journal} {\bibinfo  {journal} {Surf. Sci.}\ }\textbf {\bibinfo
  {volume} {55}},\ \bibinfo {pages} {427 } (\bibinfo {year}
  {1976})}\BibitemShut {NoStop}%
\bibitem [{\citenamefont {Bardeen}(1936)}]{Bardeen_36}%
  \BibitemOpen
  \bibfield  {author} {\bibinfo {author} {\bibfnamefont {J.}~\bibnamefont
  {Bardeen}},\ }\href {\doibase 10.1103/PhysRev.49.653} {\bibfield  {journal}
  {\bibinfo  {journal} {Phys. Rev.}\ }\textbf {\bibinfo {volume} {49}},\
  \bibinfo {pages} {653} (\bibinfo {year} {1936})}\BibitemShut {NoStop}%
\bibitem [{\citenamefont {van Himbergen}\ and\ \citenamefont
  {Silbey}(1978)}]{vanHimbergen_78}%
  \BibitemOpen
  \bibfield  {author} {\bibinfo {author} {\bibfnamefont {J.~E.}\ \bibnamefont
  {van Himbergen}}\ and\ \bibinfo {author} {\bibfnamefont {R.}~\bibnamefont
  {Silbey}},\ }\href {\doibase 10.1103/PhysRevB.18.2674} {\bibfield  {journal}
  {\bibinfo  {journal} {Phys. Rev. B}\ }\textbf {\bibinfo {volume} {18}},\
  \bibinfo {pages} {2674} (\bibinfo {year} {1978})}\BibitemShut {NoStop}%
\bibitem [{\citenamefont {Huntington}(1951)}]{Huntington_51}%
  \BibitemOpen
  \bibfield  {author} {\bibinfo {author} {\bibfnamefont {H.~B.}\ \bibnamefont
  {Huntington}},\ }\href {\doibase 10.1103/PhysRev.81.1035} {\bibfield
  {journal} {\bibinfo  {journal} {Phys. Rev.}\ }\textbf {\bibinfo {volume}
  {81}},\ \bibinfo {pages} {1035} (\bibinfo {year} {1951})}\BibitemShut
  {NoStop}%
\bibitem [{\citenamefont {Kostrobij}\ and\ \citenamefont
  {Markovych}(2015)}]{Kostrobij_15}%
  \BibitemOpen
  \bibfield  {author} {\bibinfo {author} {\bibfnamefont {P.~P.}\ \bibnamefont
  {Kostrobij}}\ and\ \bibinfo {author} {\bibfnamefont {B.~M.}\ \bibnamefont
  {Markovych}},\ }\href {\doibase 10.1103/PhysRevB.92.075441} {\bibfield
  {journal} {\bibinfo  {journal} {Phys. Rev. B}\ }\textbf {\bibinfo {volume}
  {92}},\ \bibinfo {pages} {075441} (\bibinfo {year} {2015})}\BibitemShut
  {NoStop}%
\bibitem [{\citenamefont {Han}\ and\ \citenamefont {Liu}(2009)}]{Han_09}%
  \BibitemOpen
  \bibfield  {author} {\bibinfo {author} {\bibfnamefont {Y.}~\bibnamefont
  {Han}}\ and\ \bibinfo {author} {\bibfnamefont {D.-J.}\ \bibnamefont {Liu}},\
  }\href {\doibase 10.1103/PhysRevB.80.155404} {\bibfield  {journal} {\bibinfo
  {journal} {Phys. Rev. B}\ }\textbf {\bibinfo {volume} {80}},\ \bibinfo
  {pages} {155404} (\bibinfo {year} {2009})}\BibitemShut {NoStop}%
\bibitem [{\citenamefont {Dymnikov}(2011)}]{Dymnikov_11}%
  \BibitemOpen
  \bibfield  {author} {\bibinfo {author} {\bibfnamefont {V.~D.}\ \bibnamefont
  {Dymnikov}},\ }\href {\doibase 10.1134/S106378341105009X} {\bibfield
  {journal} {\bibinfo  {journal} {Phys. Solid State}\ }\textbf {\bibinfo
  {volume} {53}},\ \bibinfo {pages} {901} (\bibinfo {year} {2011})}\BibitemShut
  {NoStop}%
\bibitem [{\citenamefont {Paskin}\ and\ \citenamefont
  {Singh}(1965)}]{Paskin_65}%
  \BibitemOpen
  \bibfield  {author} {\bibinfo {author} {\bibfnamefont {A.}~\bibnamefont
  {Paskin}}\ and\ \bibinfo {author} {\bibfnamefont {A.~D.}\ \bibnamefont
  {Singh}},\ }\href {\doibase 10.1103/PhysRev.140.A1965} {\bibfield  {journal}
  {\bibinfo  {journal} {Phys. Rev.}\ }\textbf {\bibinfo {volume} {140}},\
  \bibinfo {pages} {A1965} (\bibinfo {year} {1965})}\BibitemShut {NoStop}%
\bibitem [{\citenamefont {Smith}(1965)}]{SMITH_65}%
  \BibitemOpen
  \bibfield  {author} {\bibinfo {author} {\bibfnamefont {B.}~\bibnamefont
  {Smith}},\ }\href {\doibase https://doi.org/10.1016/0031-9163(65)90290-8}
  {\bibfield  {journal} {\bibinfo  {journal} {Phys. Lett.}\ }\textbf {\bibinfo
  {volume} {18}},\ \bibinfo {pages} {210 } (\bibinfo {year}
  {1965})}\BibitemShut {NoStop}%
\bibitem [{\citenamefont {Thompson}\ and\ \citenamefont
  {Blatt}(1963)}]{THOMPSON_63}%
  \BibitemOpen
  \bibfield  {author} {\bibinfo {author} {\bibfnamefont {C.}~\bibnamefont
  {Thompson}}\ and\ \bibinfo {author} {\bibfnamefont {J.}~\bibnamefont
  {Blatt}},\ }\href {\doibase https://doi.org/10.1016/S0375-9601(63)80003-1}
  {\bibfield  {journal} {\bibinfo  {journal} {Phys. Lett.}\ }\textbf {\bibinfo
  {volume} {5}},\ \bibinfo {pages} {6 } (\bibinfo {year} {1963})}\BibitemShut
  {NoStop}%
\bibitem [{\citenamefont {Wu}\ and\ \citenamefont {Zhang}(2008)}]{Biao_08}%
  \BibitemOpen
  \bibfield  {author} {\bibinfo {author} {\bibfnamefont {B.}~\bibnamefont
  {Wu}}\ and\ \bibinfo {author} {\bibfnamefont {Z.}~\bibnamefont {Zhang}},\
  }\href {\doibase 10.1103/PhysRevB.77.035410} {\bibfield  {journal} {\bibinfo
  {journal} {Phys. Rev. B}\ }\textbf {\bibinfo {volume} {77}},\ \bibinfo
  {pages} {035410} (\bibinfo {year} {2008})}\BibitemShut {NoStop}%
\bibitem [{\citenamefont {Sugiyama}(1960)}]{Sugiyama_60}%
  \BibitemOpen
  \bibfield  {author} {\bibinfo {author} {\bibfnamefont {A.}~\bibnamefont
  {Sugiyama}},\ }\href {\doibase 10.1143/JPSJ.15.965} {\bibfield  {journal}
  {\bibinfo  {journal} {J. Phys. Soc. Jpn.}\ }\textbf {\bibinfo {volume}
  {15}},\ \bibinfo {pages} {965} (\bibinfo {year} {1960})}\BibitemShut
  {NoStop}%
\bibitem [{\citenamefont {Wolfram{ }Research{, }Inc.}(2018)}]{Mathematica}%
  \BibitemOpen
  \bibfield  {author} {\bibinfo {author} {\bibnamefont {Wolfram{ }Research{,
  }Inc.}},\ }\href@noop {} {\emph {\bibinfo {title} {Mathematica, {V}ersion
  11.3}}}\ (\bibinfo  {publisher} {Champaign, IL},\ \bibinfo {year}
  {2018})\BibitemShut {NoStop}%
\bibitem [{\citenamefont {Czoschke}\ \emph {et~al.}(2005)\citenamefont
  {Czoschke}, \citenamefont {Hong}, \citenamefont {Basile},\ and\ \citenamefont
  {Chiang}}]{Czoschke_05}%
  \BibitemOpen
  \bibfield  {author} {\bibinfo {author} {\bibfnamefont {P.}~\bibnamefont
  {Czoschke}}, \bibinfo {author} {\bibfnamefont {H.}~\bibnamefont {Hong}},
  \bibinfo {author} {\bibfnamefont {L.}~\bibnamefont {Basile}}, \ and\ \bibinfo
  {author} {\bibfnamefont {T.-C.}\ \bibnamefont {Chiang}},\ }\href {\doibase
  10.1103/PhysRevB.72.075402} {\bibfield  {journal} {\bibinfo  {journal} {Phys.
  Rev. B}\ }\textbf {\bibinfo {volume} {72}},\ \bibinfo {pages} {075402}
  (\bibinfo {year} {2005})}\BibitemShut {NoStop}%
\bibitem [{\citenamefont {Korotun}(2015)}]{Korotun_15}%
  \BibitemOpen
  \bibfield  {author} {\bibinfo {author} {\bibfnamefont {A.~V.}\ \bibnamefont
  {Korotun}},\ }\href {\doibase 10.1134/S1063783415020213} {\bibfield
  {journal} {\bibinfo  {journal} {Phys. Solid State}\ }\textbf {\bibinfo
  {volume} {57}},\ \bibinfo {pages} {391} (\bibinfo {year} {2015})}\BibitemShut
  {NoStop}%
\bibitem [{\citenamefont {Kubo}(1962)}]{Kubo_62}%
  \BibitemOpen
  \bibfield  {author} {\bibinfo {author} {\bibfnamefont {R.}~\bibnamefont
  {Kubo}},\ }\href {\doibase 10.1143/JPSJ.17.975} {\bibfield  {journal}
  {\bibinfo  {journal} {J. Phys. Soc. Jpn.}\ }\textbf {\bibinfo {volume}
  {17}},\ \bibinfo {pages} {975} (\bibinfo {year} {1962})}\BibitemShut
  {NoStop}%
\bibitem [{\citenamefont {Waber}\ \emph {et~al.}(1972)\citenamefont {Waber},
  \citenamefont {Kennard},\ and\ \citenamefont {Tsui}}]{waber_72}%
  \BibitemOpen
  \bibfield  {author} {\bibinfo {author} {\bibfnamefont {J.}~\bibnamefont
  {Waber}}, \bibinfo {author} {\bibfnamefont {E.}~\bibnamefont {Kennard}}, \
  and\ \bibinfo {author} {\bibfnamefont {Y.-P.}\ \bibnamefont {Tsui}},\ }\href
  {\doibase 10.1051/jphyscol:1972315} {\bibfield  {journal} {\bibinfo
  {journal} {J. Phys. Colloques}\ }\textbf {\bibinfo {volume} {33}},\ \bibinfo
  {pages} {C3} (\bibinfo {year} {1972})}\BibitemShut {NoStop}%
\end{thebibliography}
%\input{metalsurface.bib} 
%\end{thebibliography}

\newpage

\end{document}